\documentclass[aps,prb,groupedaddress,showpacs,preprintnumbers,amsmath,amssymb,twocolumn,floats]{revtex4}

\usepackage{graphicx}
\usepackage[dvipdfm]{hyperref}
\usepackage{booktabs}
\usepackage{float}
\usepackage[]{sidecap}
\usepackage{CJK}
\usepackage{txfonts}
\usepackage{multirow}

\begin{document}
\begin{CJK*}{UTF8}{gbsn}

\title{The design of a high flux VUV beamline for low energy photons}

\author{ Jiajia Wang $^1$ }
\author{Mao Ye  $^2$ }
\author{ Tan Shi $^1$ }
\author{Rui Chang  $^2$ }
\author{Shan Qiao  $^2$ $^3$}\email{qiaoshan@mail.sim.ac.cn}

\affiliation{1. Physics Department, Laboratory of advanced Materials and Surface Physics Laboratory , Fudan University, Shanghai, 200438, China}
\affiliation{2.Statekey laboratory of functional materials for informatics, Shanghai Institute of Microsystem and Information Technology, Chinese Academy of Sciences, 865 Changning Road, Shanghai, 200050, China }
\affiliation{3.School of physical science and technology, ShanghaiTec University, 319 Yueyang Road,Shanghai 200031, China}

\date{\today}


\begin{abstract}

A VUV beamline at SSRF for ARPES measurements are designed. To increase the resolution and bulk sensitivity, the photon energy as low as 7 eV is desired. Because the reflectivity for p-polarized photons strongly decreases when the photon energy is below 30 eV, the design of high flux beamline for low energy VUV photons is a challenge. This work shows a variable including angle VLPGM with varied grating depth (VGD) which can achieve both high resolution and high flux with broad energy coverage.
\end{abstract}


\maketitle
\end{CJK*}

\section{Introduction}

Photoelectron spectroscopy (PES) is one of the most powerful methods that can directly investigate the electronic structures of condensed matter. The angle-resolved PES (ARPES) further provide the momentum information of electrons in the solids, which makes the direct visualization of electronic energy band possible. However, the surface-sensitivity of PES limits the observation of electronic structures near  the surface, due to the relatively short mean-free-path of photoelectrons ~\cite{lab1}. In order to obtain higher bulk-sensitivity, as well as higher resolution for energy and momentum, low energy photons with energy bellow 20 eV is one of the best choices. On the other hand, PES experiments with higher photon energy, up to several hundred electron volts (eV) would also be necessary for the study of core level electronic states. Furthermore, the polarization of the incident photons provides a new degree of freedom in the APRES measurements, which enable us to selectively probe electrons from different orbitals with different symmetries. Thus, a beamline, which can cover low photon energy down to 7 eV because of the availability of laser below this energy, and up to several hundred electronvolts with various polarization and sufficient photon flux, is highly hoped for the study of condensed matter physics by means of APRES.
HRPES(High Resolution PhotoElectron Spectroscopy) beamline is constructed in SSRF, Shanghai Synchrotron Radiation Facility. The purpose of this beamline is for high resolution ARPES. The energy range of this beamline is from 7 to 791 eV. The source of this beamline is an APPLE-Knot undulator, which is a permanent magnetic one realizing the character of Knot undulator ~\cite{lab2} and can strongly suppress the heat load on the optics axis and supply photons with arbitrary polarizations. The targets of the design are:

1)High energy resolution and larger than 20000 resolving power is preferred.

2)A small spot size on sample because the efficiency, and the signal-to-noise ratio of the electron analyzer is strongly depended on it.

3)Not very low reflectivity for vertical polarized photons. For low energy VUV photons, the reflectivity of p-polarized photons decreased strongly compared with s-polarized ones, special care is needed to overcome this problem.

\section {Beamline design}

The geometry limitation of the beamline is as following. The undulator source is 1.3 meter above the floor of the experimental hall. The outsider of the end front is 18.0965 meter from the source which is the center of the undulator. The source sample distance is 40.3 meter and the sample level is 1 meter above the floor.

The effective electron beam size $\sigma_{r0}$ and divergence  $\sigma_{r0}^\prime$ of the storage ring of SSRF in vertical (horizontal) direction are 8.36 $\mu$m (143.2 $\mu$m) and 4.18 $\mu$rad (34  $\mu$rad), respectively. The total length of the undulator is 4.5 meter. Then from the consideration of spatial coherence (diffraction limit), the rms of size $\sigma_r$ and divergence angle $\sigma_r^\prime$ of the source are ~\cite{lab3}
\begin{eqnarray}
\sigma_r = \sqrt{\sigma_{r0}^2 + \frac{1}{2\pi^2}\lambda L}
\end{eqnarray}
\begin{eqnarray}
\sigma_r^\prime = \sqrt{\sigma_{r0}^{\prime2} + \frac{\lambda}{2L}}
\end{eqnarray}
where $\lambda$ and $L$ are the wavelength of the photon and the total length of the undulator, respectively. We set the distance between source and exit slit as 35 m£¬source grating distance $r$ as 22 m and grating exit slit distance $r^\prime$ as 13 m.

\begin{figure}
\includegraphics[width=8cm]{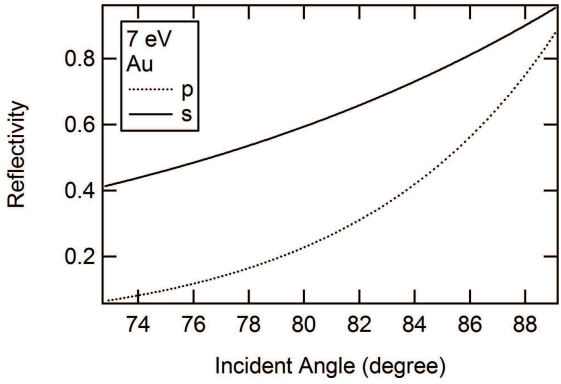}
\caption{  The reflectivity of s- and p-polarized 7 eV photons off Au surface at different incident angles calculated by the XOP program developed by ESRF.
 }\label{fig1}
\end{figure}

For VUV beamline cover photon energy as low as 7 eV, some special care should be taken. For 7 eV photons, the source size and divergence angle estimated from the above equations are 201 $\mu$m (247 $\mu$m) and 0.14 mrad (0.144 mrad) in vertical (horizontal) direction. To accept almost all photons, usually a $4\sigma_r^\prime$ acceptance angle is needed and which is about 0.6 mrad. Because of the acceptance angle as large as 0.6 mrad and the limited grating length of about 130 mm, the incident angle $\alpha$, which is the angle between the photon beam and the normal of grating, should be less than 84.17 degrees. On the other hand, from the reflectivity of 7 eV photons (Fig.~\ref{fig1}), we can see the reflectivity of p-polarized photons decreases strongly when the incident angle reduced£¬so if we hope higher flux for low energy vertical polarized photons, the beamline cannot select NIM (normal incident monochromator) type which is widely be used for low energy VUV beamline, and we should select $\alpha$ as possible as close to 84.17 degrees for 7 eV photons.

There are four popular designs of monochromator for high resolution VUV beamline, dragon type, PGM (Plane-Grating Monochromator) with collimated light, VLPGM (Varied Line-space Plane-Grating Monochromator) with and without pre-focus mirror. The advantage of dragon ~\cite{lab4} type beamline is the simplest structure that only one grating is used which results in high efficiency. The disadvantage of this type is that the exit slit need to move to change the exit arm length $r^\prime$ when scan photon energy, which results in the broadening of spot on sample which is against the targets of our design. The entrance and exit light beams are parallel for PGM with collimated light ~\cite{lab5,lab6} which results in the freedom to change $C_{ff}$. However, this design can only work for grating with constant line-space and a refocus mirror is needed which will decrease the photon flux which again disagrees our target. For VLPGM with pre-focus mirror, there are two configurations. If the mirror make an image source behind the grating with $r \approx -r^\prime$ ~\cite{lab7}, the accurate focus can be achieved with fix exit length arm with constant including angle at two photon energies and the defocus at other photon energies can be small enough. The advantage of this configuration is that the scan of photon energy can be performed with the only rotation of grating. The second configuration~\cite{lab8} is to set an image source made by the pre-mirror at a certain distance behind the grating, then the accurate focus can be achieved in two photon energies with two selectable different including angles. Although in principle, the accurate focus at different energies needs different including angles, Amemiya and Ohta showed that the good enough performance could be achieved with the fixed two including angles for the whole energy range and the energy scan could be done with only the rotation of grating. For our beamline, because the divergence of photon source decreases with the increase of photon energy, a larger incidence angle is needed for photons with higher energy to increase the reflectivity and variable including angle are preferred. Although the second configuration of VLPGM with pre-focus can satisfy this requirement, because the pre-focus mirror may result in some aberrations, we choose VLPGM without pre-focus as the fundamental of our design. In our design, the almost same shine length on grating can be achieved, which means that the maximum incident angles or say maximum reflectivity have been achieved for whole energy range. For G1, the shine length changes from 119 mm at 7 eV to 125 mm at 104 eV. Suppose the line density of the VLPGM is

\begin{eqnarray}
k = k_0(1+2b_2w+3b_3w^2)
\end{eqnarray}
where the positive direction of $w$ is along the photon transfer direction, then the focus condition for -1 order diffraction is

\begin{eqnarray}
\frac{cos^2\alpha}{r} + \frac{cos^2\beta}{r^\prime} - 2b_2 k_0\lambda=0
\end{eqnarray}
When the practice distance between source and grating is different from theoretical value, the focus condition can be remedied by the change of $\alpha$  and $\beta$  achieved by the rotations of pre-mirror and grating. This is another big advantage of our design.

In our design, three gratings are used and the including angle changes from 158 to 174.4 degree. G1 with line density of 190 covers energy from 7 to 104 eV, G2 with line density of 620 covers energy from 17 to 244 eV, and G3 with line density of 2000 covers energy from 55 to 791 eV. All the gratings are coated with Au.

\begin{figure}
\includegraphics[width=8cm]{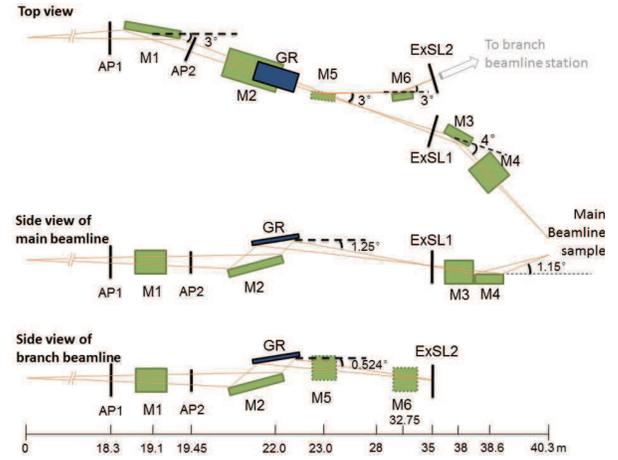}
\caption{(Color online)  The layout of the beamline.
 }\label{fig2}
\end{figure}
The optics layout of the beamline is shown in Fig.~\ref{fig2}. White light aperture AP1 and AP2 are set at 18.3 and 19.45 meter to control the acceptance angle of the beamline and to cut the diffuse scattered light from the M1. Wire beam position monitor WBPM 1 and WBPM 2 are set just before AP1 and AP2 for the observation of position and emission angle of white light beam. A spherical mirror M1 at 19.1 meter is used to focus photon beam horizontally on position at 28 m and deflect the beam horizontally for 3 degree. The magnification of this focusing is 0.466. The grazing angle of M1 is chosen as 1.5 degree and then the length of M1 should be 450 mm with a 438 mm effective length to collect all the photons with 0.6 mrad opening angle. To decrease the mirror length, the mirror is set as near as possible to the shield wall. The grating is at 22 meter from source and the photon beam leaving grating go away from horizontal direction with an angle of 1.25 and 0.524 degree towards the floor of experimental hall for main and branch beamlines, respectively. The exit slits are at 35 meter from source. The vertical magnification of grating is
\begin{eqnarray}
M_G = \frac{cos\alpha/r}{cos\beta/r^\prime}
\end{eqnarray}

\begin{figure}
\includegraphics[width=8cm]{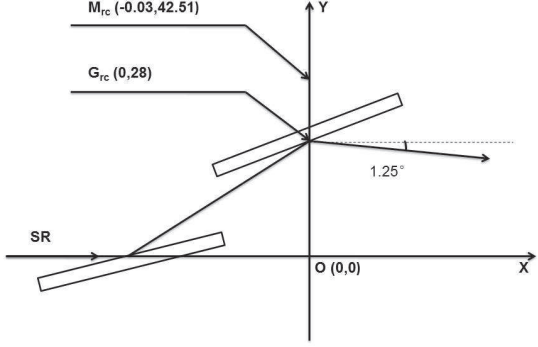}
\caption{  The geometry of M2 rotation mechanism.
 }\label{fig3}
\end{figure}
For 7 eV photons, $M_G$ is 0.221. An elliptical cylindrical mirror at 38 m focuses the photon beam to the sample position at 40.3 m horizontally. Another one at 38.6 m focuses the photon beam from exit slit to sample position vertically. To keep the fixed direction of exit photons, the M2 mirror needs a rotation with a horizontal movement. The complex movement can be achieved by the off-axis rotation of M2. The weakness of off-axis rotation is the very long M2 and the length of M2 should be as long as 392 mm. The distance between entrance light and the rotation center of grating is set as 28 mm. The best performance is achieved when we set the rotation axis of M2 at (-0.03 mm, 14.51 mm) referring to the grating center as shown in Fig.~\ref{fig3} and the movement of light on the grating surface is smaller than 16.4 $\mu$m during energy scan. A M5 mirror at 23 m is used to switch the beam between main and branch beamlines. A M6 at 32.75 meter is used to focus the beam horizontally to the exit slit at 35 m of the branch beamline. The parameters for all optics elements are shown in Table~\ref{tab1} and Table~\ref{tab2}.

\begin{table*}[htbp]
\centering
\caption{\label{tab1}  The parameters of optical elements.}
\begin{tabular*}{170mm}{@{\extracolsep{\fill}}ccccccc}
\toprule  & Figure  & Distance(m) & Def.Ang(degree) & R/$\rho$(m/m) & Effective Size(mm)  &Size(mm)\\ \hline
M1 & spherical & 19.1 & 3 & 464.067 & 438$\times$12 & 450$\times40$$\times50$\\
M2& plane & 21.9261-21.6319 & 20.75-4.35 & & 382$\times$8 & 392$\times40$$\times50$\\
G1 & VLSPG & 22.0051-22.0011 & 158-174.4 & & 125$\times$8 & 137$\times30$$\times50$\\
G2 & VLSPG & 22.0051-22.0011 & 158-174.4&  & 127$\times$8 & 137$\times30$$\times50$\\
G3 & VLSPG & 22.0051-22.0011 & 158-174.4&  & 73$\times$8 & 85$\times30$$\times50$\\
M3 & Ellip. Cyl & 38 & 4.0& 6.151/0.1674 & 369$\times$9 & 380$\times30$$\times60$\\
M4 & Ellip. Cyl & 38.6 & 2.4& 3.151/0.0576 & 490$\times$14 & 500$\times35$$\times60$\\
M5 & plan & 23 & 3.0&  & 246$\times$33 & 260$\times50$$\times60$\\
M6 & Ellip. Cyl & 32.75& 3.0& 3.501/0.0856 & 234$\times$7 & 250$\times30$$\times60$\\
\bottomrule
\end{tabular*}%
\end{table*}

\begin{table*}
\centering
\caption{ \label{tab2}  Grating parameters.}
\begin{tabular*}{170mm}{@{\extracolsep{\fill}}ccccccc}
\toprule   Grating  & K0 $mm^{-1}$ & Energy range & b2 10$^{-5}$$mm^{-1}$ & b3 10$^{-9}$mm$^{-2}$  &-$\alpha$($^0$)&$\beta$($^0$)\\ \hline
G1 & 190 & 7-104 & 9.466 & 6.431& 84.06-88.53 & 73.94-85.87\\
G2 & 620 & 17-244 & 8.220& 6.032 & 85.81-89.04 & 72.19-85.35\\
G3 & 2000 & 55-791 & 8.229& 6.035 & 85.79-89.04 & 72.21-85.36\\
\bottomrule
\end{tabular*}%
\end{table*}

\begin{figure}
\includegraphics[width=8cm]{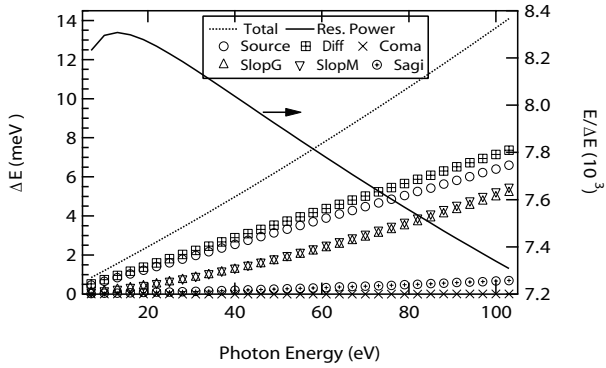}
\caption{ The resolution of 190 mm$^{-1}$ G1.
 }\label{fig4}
\end{figure}

\begin{figure}
\includegraphics[width=8cm]{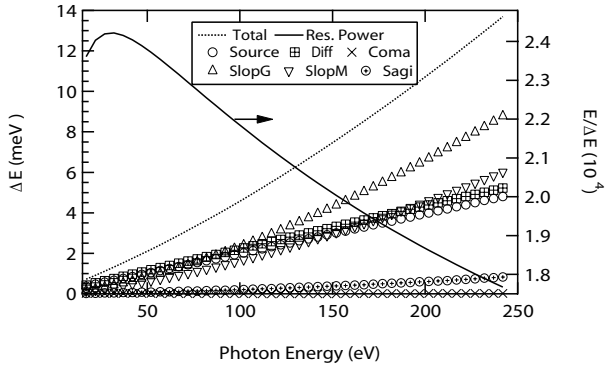}
\caption{  The resolution of 620 mm$^{-1}$ G2.
 }\label{fig5}
\end{figure}

\begin{figure}
\includegraphics[width=8cm]{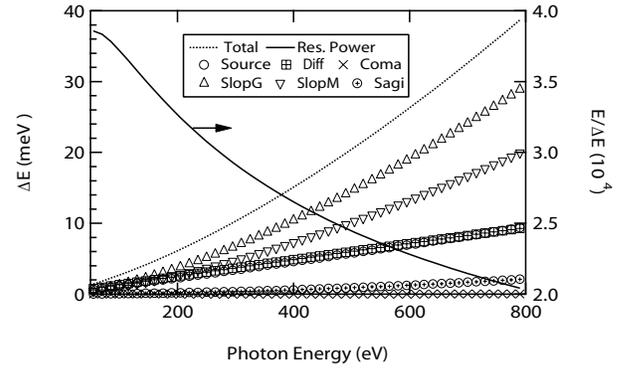}
\caption{  The resolution of 2000 mm$^{-1}$ G3.
 }\label{fig6}
\end{figure}

The total vertical magnification for 7eV photon in main beamline is
\begin{eqnarray}
M_V = 0.221\times\frac{1.7}{3.6} = 0.104
\end{eqnarray}
The total horizontal magnification is
\begin{eqnarray}
M_H = \frac{8.9}{19.1}\times\frac{2.3}{10} = 0.107
\end{eqnarray}
The FWHM of the beam size on sample can be estimated from above magnifications as 49.1 and 62.1 $\mu$m along vertical and horizontal directions. The energy resolutions of gratings related with source size, diffraction limitation, slop error of M2, slop error of grating, coma and sagital focus can be calculated with the following equations.

\begin{eqnarray}
(\frac{\Delta E}{E})_{so} = (\frac{\Delta E}{E})_{slit} = \frac{2.35\sigma_v cos\alpha}{\lambda k_0 r}
\end{eqnarray}

\begin{eqnarray}
(\frac{\Delta E}{E})_{diff} = \frac{1}{k_0 l} = \frac{cos\alpha}{k_0 2.35 \sigma_v^\prime r}
\end{eqnarray}

\begin{eqnarray}
(\frac{\Delta E}{E})_{sm} = \frac{2.35 \times 2\sigma_m}{\lambda k_0}cos\alpha
\end{eqnarray}

\begin{eqnarray}
(\frac{\Delta E}{E})_{sg} = \frac{2.35\sigma_g}{\lambda k_0}(cos\alpha + cos\beta)
\end{eqnarray}

\begin{eqnarray}
(\frac{\Delta E}{E})_{coma} = (\frac{2.35 \times 0.5\sigma_v^\prime r}{cos\alpha})^2 \frac{3F_{30}}{2\lambda k_0}
\end{eqnarray}

\begin{eqnarray}
(\frac{\Delta E}{E})_{sag} = \frac{(2.35 \times 0.5\sigma_h^\prime r)^2}{2\lambda k_0} (F_{12} + sin\beta F_{02}^2)
\end{eqnarray}
where $\sigma_v$, $l$, $\sigma_v^\prime$, $\sigma_m$, $\sigma_g$ and $\sigma_h^\prime$ are vertical source size, shine length, vertical source divergence, mirror slop error, grating slop error, horizontal source divergence and

\begin{eqnarray}
F_{30} = \frac{sin\alpha cos^2\alpha}{r^2} + \frac{sin\beta cos^2\beta}{r^{\prime2}}-2\lambda k_0 b_3
\end{eqnarray}
\begin{eqnarray}
F_{12} = \frac{sin\alpha}{r^2} + \frac{sin\beta}{r^{\prime2}}, F_{02} = \frac{1}{r} + \frac{1}{r^\prime}
\end{eqnarray}
here $F_{30}$ is set zero at the midpoint of the energy range by choosing a suitable $b_3$. The results are shown in Fig.~\ref{fig4}-Fig.~\ref{fig6}, respectively. In the calculations, the slope error of grating is set as 0.5 $\mu$rad and the slope error of M2 is set as 1 $\mu$rad after consider the effect of heat load. The G1 is for high flux at low energy range, and G2 and G3 are for high energy resolution. We can see the energy resolving power is above 20000 for photon energy above 17 eV. The energy resolution is below 2 meV for photon energy lower than 17 eV and good enough for ARPES experiments. For G1, the diffraction limitation and source size are the main contribution to the energy resolution because of the low line density. For G2 and G3, the slop error of grating is the main contribution. For all gratings, the coma and sagittal focus are negligible.

\begin{figure}
\includegraphics[width=8cm]{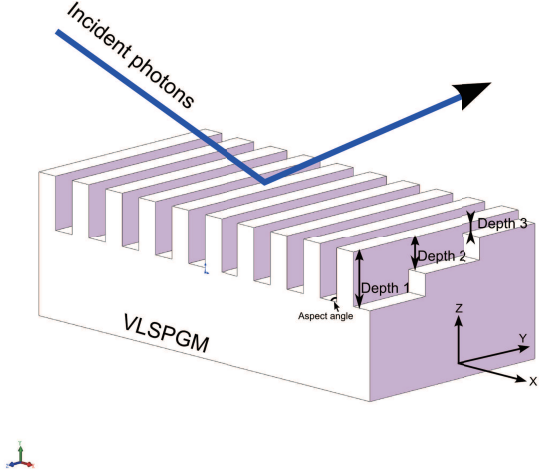}
\caption{(Color online)  Schematic design for the VLSPGM with VGD profile.
 }\label{fig7}
\end{figure}

\begin{figure}
\includegraphics[width=8cm]{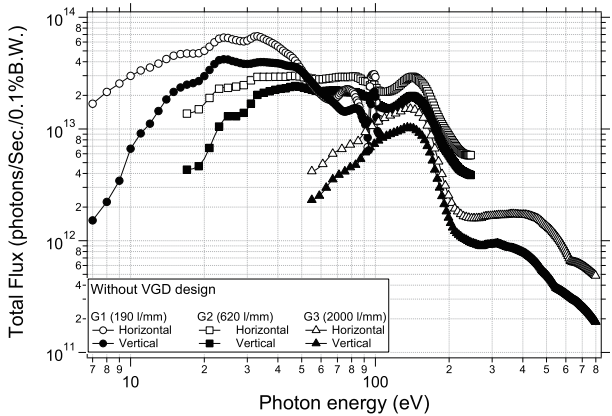}
\caption{ The flux on sample without VGD structure.
 }\label{fig8}
\end{figure}

\begin{figure}
\includegraphics[width=8cm]{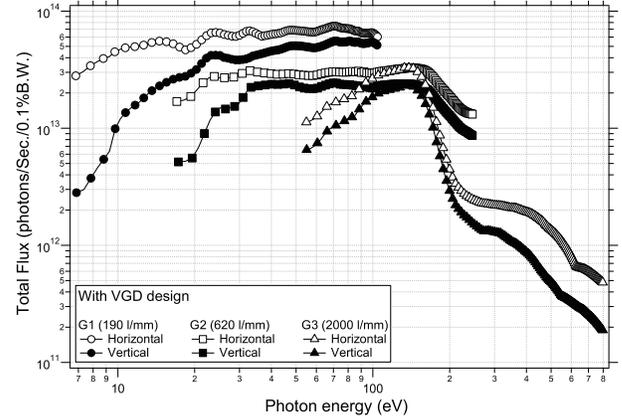}
\caption{  The flux at sample position.
 }\label{fig9}
\end{figure}

\begin{figure}
\includegraphics[width=8cm]{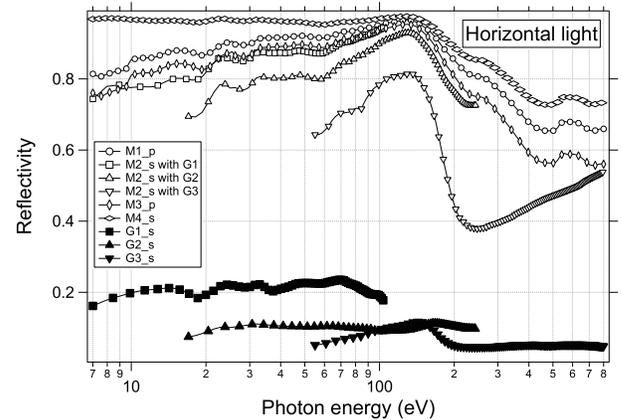}
\caption{  The efficiencies of every element for horizontal polarized photons.
 }\label{fig10}
\end{figure}

\begin{figure}
\includegraphics[width=8cm]{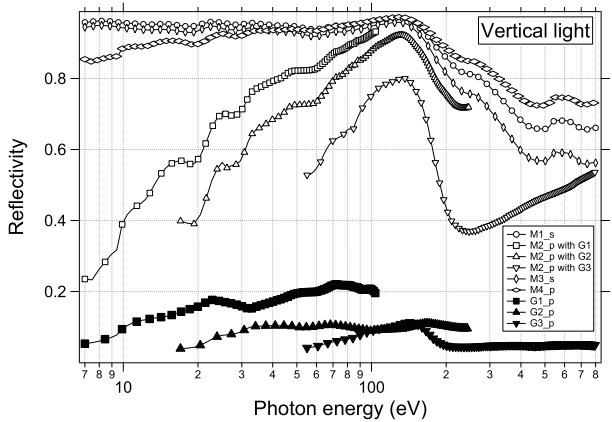}
\caption{  The efficiencies of every element for vertical polarized photons.
 }\label{fig11}
\end{figure}

In grating design, the grating groove depth can only be optimized for certain photon energy. In order to achieve good efficiency of the gratings as high as possible for different photon energies, we chose varied-groove-depth (VGD) type gratings. For simplicity, only three kind groove depths are calculated for each grating and optimized by the code REFLECT~\cite{lab9,lab10}. In practice, continually changed VGD will be used.  The areas with different groove depths are located at different positions of each grating along the direction perpendicular to the optical path as shown in Fig.~\ref{fig7}. Thus by shifting the illuminated area on the grating along the y-direction indicated in Fig.~\ref{fig7}, the highest efficiency can be easily selected for different photon energies. The parameters chosen for each grating are shown in Table~\ref{tab3}. The fluxes at sample position without and with VGD structure are shown in Fig.~\ref{fig8} and Fig.~\ref{fig9}. They were calculated by the code REFLECT. We can see the VGD structure can increase the flux by 67\% and 84\% respectively for horizontal and vertical polarized photons at 7 eV and VGD can make the flux more flat at high energy side. The reflectivity of mirrors and VGD gratings  is  shown in Fig.~\ref{fig10} and Fig.~\ref{fig11} for horizontal and vertical polarized photons, respectively.
\begin{table*}
\centering
\caption{ \label{tab3}  Parameters for VGD gratings.}
\begin{tabular*}{170mm}{@{\extracolsep{\fill}}cccccccc}
\toprule    & Depth 1& Depth 2& Depth 3 & Coating   &Coating &Aspect angle&Groove width \\
            & (nm)& (nm)& (nm) &  material  &thickness (nm)&($^0$)& / Spacing ratio\\ \hline
G1 & 140 & 100 & 70 & Au& 60 & 90&0.65\\
G2 & 45 & 35 & 20& Au & 25 & 90&0.65\\
G3 & 12 & 10 & 7&Au & 40 & 90&0.65\\
\bottomrule
\end{tabular*}%
\end{table*}

\begin{figure}
\includegraphics[width=7cm]{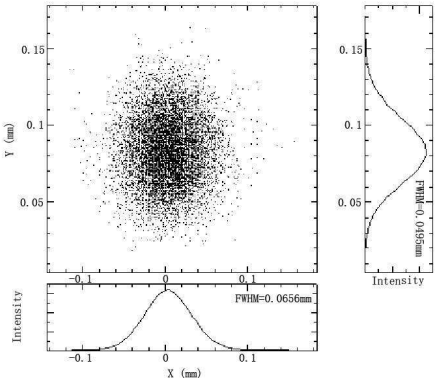}
\caption{  Ray tracing result of spot size on sample by RAY program.
 }\label{fig12}
\end{figure}

\begin{figure}
\includegraphics[width=8cm]{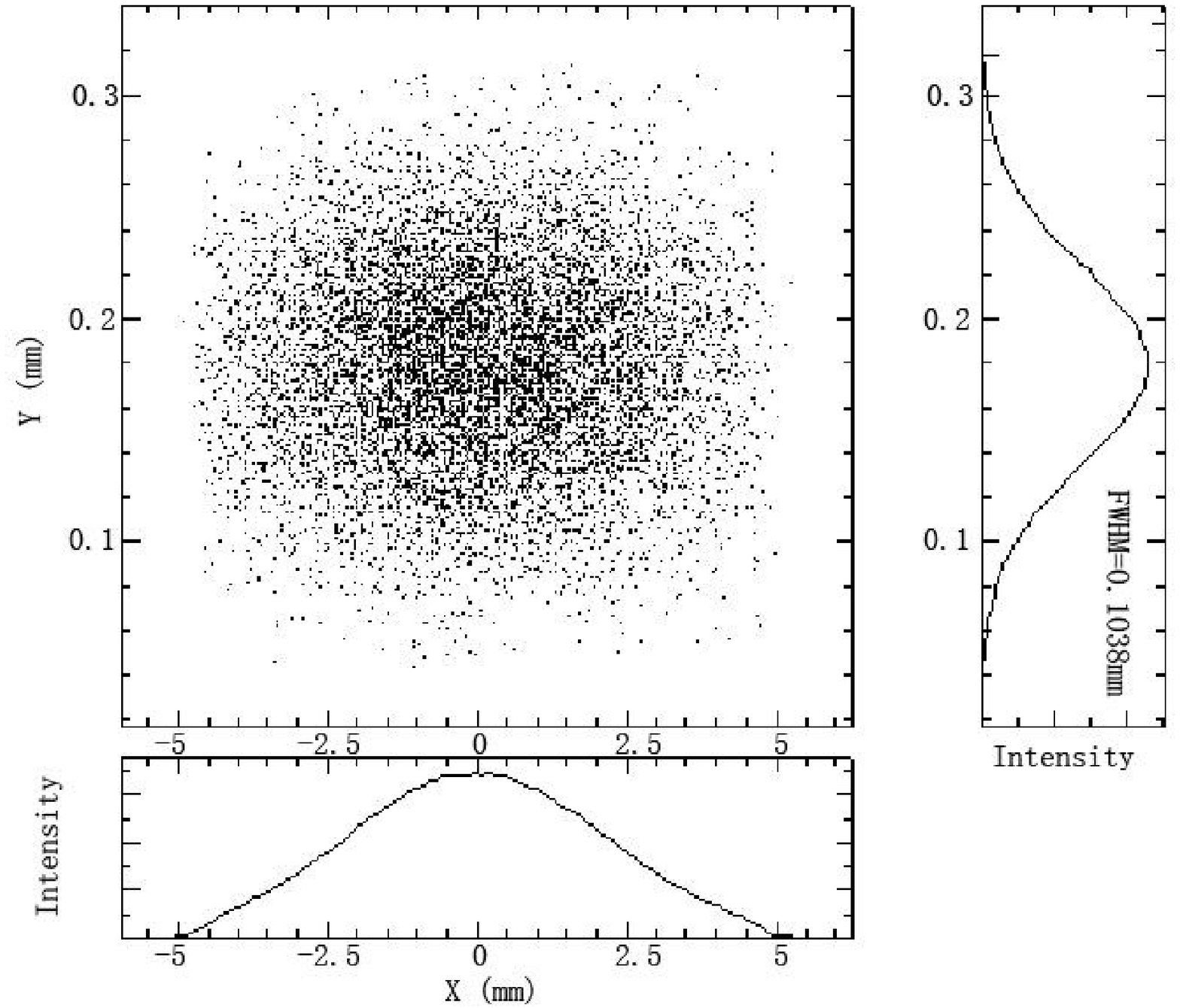}
\caption{  The beam size on the slit for 7 eV photons.
 }\label{fig13}
\end{figure}
Ray tracing was done by the code RAY ~\cite{lab11} for 7eV photons, which possesses the largest divergence. In the calculations, the slope errors of grating and mirrors are not included. The spot size on sample without exit slit is shown in Fig.~\ref{fig12}. The FWHM of spot sizes on sample are 49.5 and 65.6 $\mu$m along vertical and horizontal directions respectively. The beam size on exit slit is shown in Fig.~\ref{fig13} for 7 eV photons and the FWHM of beam size in vertical direction is 103.8 $\mu$m.

\section{Summary}
We have shown the design of the HRPES beamline at SSRF which can supply photons with energy from 7 to 791 eV with arbitrary polarization. The energy resolution power is larger than 20000 when photon energy higher than 17 eV. The energy resolution of photons with energy below 17 eV is less than 2 meV, which is good enough for ARPES measurements. Meanwhile the G1 supplies higher flux for low energy photons and a photons flux greater than 10$^{12}$ photons/s 0.1\% BW can be achieved for photons in the energy range from 7 to 300 eV with an optimized VGD design of gratings. The spot sizes on sample can achieve 49.5 times 65.6 $\mu$m which is good enough for ARPES measurements.

\end{document}